\documentclass[
superscriptaddress,
amsmath,
amssymb,
twocolumn,
prd,
showpacs
floatfix,
]{revtex4-2}

\usepackage{verbatim}
\usepackage{graphicx}          
\usepackage{bm}                
\usepackage{hyperref}          
\usepackage[dvipsnames]{xcolor}

\begin{document}

\title{Implementation of residual nucleus de-excitations associated with proton decays in $^{12} {\rm C}$ based on the GENIE generator and TALYS code}

\author{Hang Hu}
\email{huhang3@mail2.sysu.edu.cn}
\affiliation{School of Physics, Sun Yat-Sen University, Guangzhou, 510275, China}
\author{Wan-Lei Guo}%
\email{guowl@ihep.ac.cn (corresponding author)}
\affiliation{Institute of High Energy Physics, Chinese Academy of Sciences,
Beijing 100049, China}
\author{Jun Su}
\email{sujun3@mail.sysu.edu.cn}
\affiliation{Sino-French Institute of Nuclear Engineering and Technology, Sun Yat-Sen University, Zhuhai, 519000, China}
\author{Wei Wang}
\email{wangw223@mail.sysu.edu.cn}
\affiliation{School of Physics, Sun Yat-Sen University, Guangzhou, 510275, China}
\affiliation{Sino-French Institute of Nuclear Engineering and Technology, Sun Yat-Sen University, Zhuhai, 519000, China}
\author{Cenxi Yuan}
\email{yuancx@mail.sysu.edu.cn}
\affiliation{Sino-French Institute of Nuclear Engineering and Technology, Sun Yat-Sen University, Zhuhai, 519000, China}

\date{\today}

\begin{abstract}
We implement the de-excitation processes of residual nuclei associated with proton decays in $^{12} {\rm C}$ based on the GENIE generator and TALYS code. To derive the reasonable excitation energy spectra of residual nuclei $^{11} {\rm B}$, $^{10} {\rm B}$ and $^{10} {\rm Be}$, the default GENIE nucleon decay generator is modified in terms of the Spectral Function nuclear model. Then we use the TALYS code to estimate the de-excitation processes of residual nuclei. The TALYS calculation can partly account for the experimental data.


\end{abstract}

\maketitle


\section{\label{sec1} Introduction}

The proton decay is a generic prediction in the Grand Unified Theories (GUTs) \cite{Georgi:1974sy, Nath:2006ut}, since these theories put quarks and leptons into the same multiplet of the GUT gauge group. The GUTs can naturally unify the strong, weak, and electromagnetic interactions into a single underlying force at a very high energy scale. The proton decay is a crucial test of the GUTs.  Although many searches have been performed over multiple decades, no experimental evidence to date for proton decay has been found \cite{PDG}. The Super-Kamiokande experiment gives the best limits to the proton lifetime for most proton decay modes. In the future, three types of detectors will continue to look for the proton decay \cite{Babu}: Water Cherenkov detectors \cite{Hyper-K}, liquid scintillator detectors \cite{JUNO,JUNO_PPNP} and liquid Argon TPC detectors \cite{DUNE}. Here we focus on the large liquid scintillator detectors for future proton decay searches.

The liquid scintillator (LS) as the detection medium in the past neutrino experiments has achieved great successes \cite{Cowan:1992xc,KamLAND, DYB}. Now the next generation large LS detectors JUNO \cite{JUNO,JUNO_PPNP} and LENA  \cite{LENA} are constructing in China and proposed in Europe, respectively. As a calorimeter, the LS detector has some specific advantages, such as its good energy resolution and a very low energy threshold, in searching for the proton decay \cite{JUNO,LENA,Borexino:2003igu,Undagoitia:2005uu,KamLAND:2005pen,KamLAND:2015pvi}. The initial decayed proton may be the free proton from H, or the bound proton in $^{12} {\rm C}$. The low energy threshold ensures that the LS detector can identify a 2.2 MeV $\gamma$-ray from the neutron captured by a proton. It can be used to separate proton decay signals from atmospheric neutrino backgrounds. Due to the excellent energy resolution, the LS experiment can measure the visible energy of daughter particles from the proton decay very well, which depends on the initial energy and momentum of the decayed proton. For the bound proton decay, one should consider the following nuclear effects \cite{Super-Kamiokande:2014otb}, Fermi motion, binding energy, and nucleon-nucleon (NN) correlation, to describe the initial state of the decayed proton. On the other hand, the final state interaction (FSI) and de-excitation processes of the residual nucleus associated with the proton decay may also change the visible energy and simultaneously produce secondary neutrons. These secondary neutrons will reduce the ability to suppress the atmospheric neutrino background through the neutron identification. To model these nuclear effects, a proton decay Monte Carlo (MC) generator is necessary for LS experiments to search for the proton decay.

GENIE is a neutrino MC generator and widely used by many neutrino experiments \cite{GENIE, GENIE:2021npt}. The recent versions of GENIE can simulate more than 60 kinds of nucleon decay modes. In this paper, we used GENIE version 3.0.2 with the G18\_02a\_00\_000 model set \cite{GENIE:2021npt}. In this configuration of GENIE, the Bodek-Ritchie nuclear model \cite{Bodek:1981wr} with short-range NN correlation is used for all processes, since it is simple and applicable across a broad range of target atoms and neutrino energies. For the proton decay in $^{12} {\rm C}$, the initial-state proton energy is limited to be a very narrow range of $910.0 \, {\rm MeV} \leq E_p \leq 922.4$ MeV, which does not match with the measured $s$-shell and $p$-shell binding energies from the $^{12} {\rm C}(e,e' p) ^{11} {\rm B}$ reaction \cite{Nakamura:1976mb}. On the other hand, the residual nucleus ${^{11} {\rm B}}$ in GENIE would be left in the ground state. Namely, the GENIE generator does not consider the de-excitation of the residual nucleus. In fact, the Super-Kamiokande experiment has used $\gamma$-ray emissions from the de-excitation processes of ${^{15} {\rm N}}$ \cite{Ejiri:1993rh} to search for $p \rightarrow \bar{\nu} K^+$ in ${^{16} {\rm O}}$ \cite{Super-Kamiokande:2014otb}. For LS experiments, the neutron emission of residual nucleus is more crucial than other particle emissions. The emitted neutrons from the $s$-hole state of $^{11} {\rm C}$ have been used by the KamLAND experiment to identify the neutron invisible decay mode in $^{12} {\rm C}$ \cite{Kamyshkov:2002wp,KamLAND:2005pen}.

In this paper, we modify the GENIE nucleon decay generator in terms of the Spectral Function (SF) nuclear model \cite{Benhar:2005dj} to account for the nuclear effects, and then use the TALYS code \cite{TALYS} to implement the de-excitation processes of the residual nucleus. The JUNO has adopted this work as a part of their simulations to give a preliminary sensitivity to  $p \rightarrow \bar{\nu} K^+$ \cite{JUNO_PPNP}. This paper is organized as follows. In Sec. II, we present the GENIE modifications, and then show the energy and momentum distributions of bound protons in the default and modified GENIE. Sec. III describes the de-excitation processes of residual nuclei $^{11} {\rm B}$, $^{10} {\rm B}$ and $^{10} {\rm Be}$ based on the TALYS software and their excitation energy spectra. In Sec. IV, we compare the predicted results with the quasifree $^{12} {\rm C} (p,2p) ^{11} {\rm B}$ data. Finally, a summary will be given in Sec. V.

\section{\label{sec2} The bound protons in $^{12} {\rm C}$}

\begin{figure}
\includegraphics[scale=0.4]{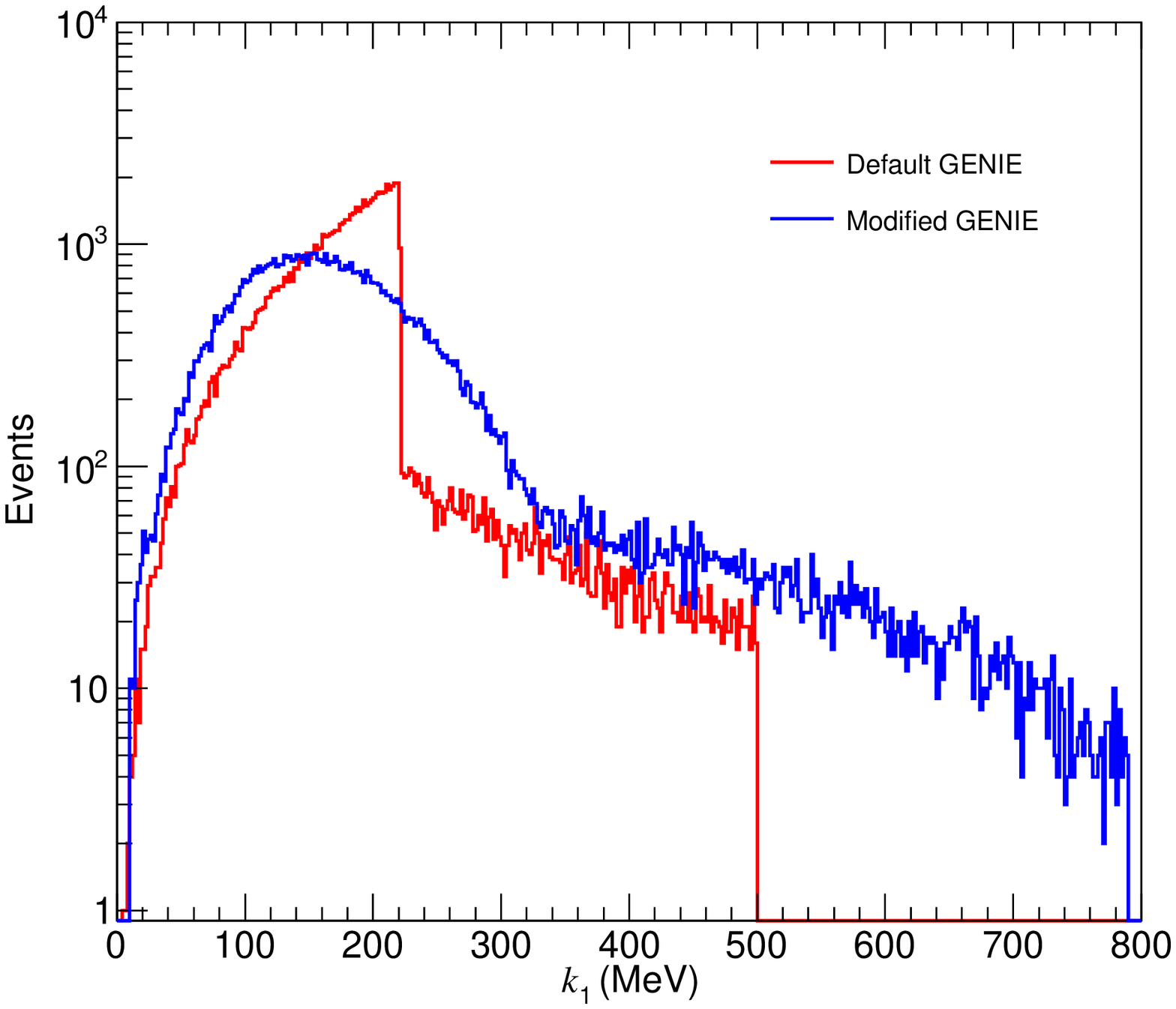}
\includegraphics[scale=0.4]{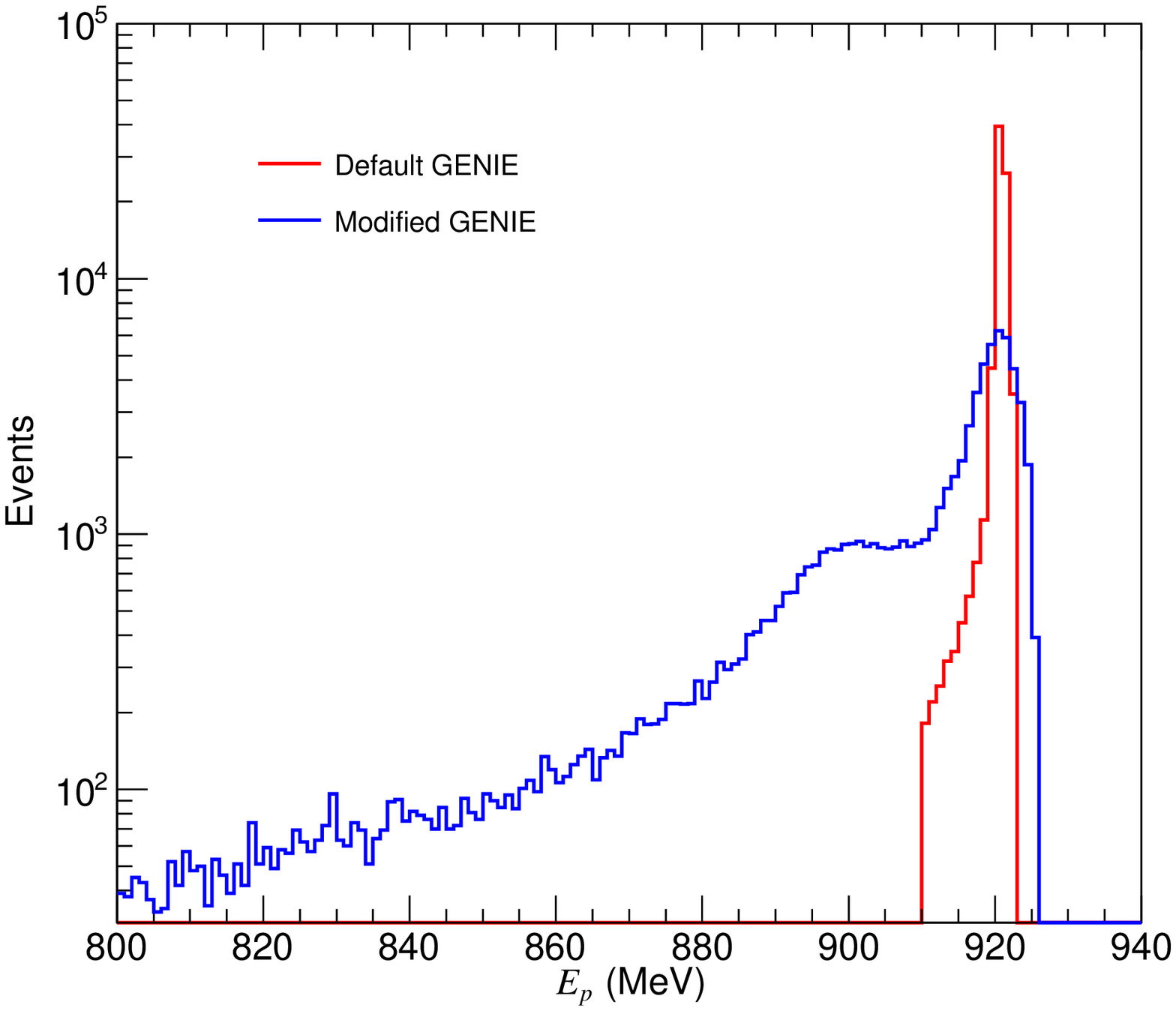}
\caption{Momentum (top) and energy (bottom) distributions of bound protons from $^{12} {\rm C}$ in the default (red) and modified (blue) GENIE.}
\label{fig1}
\end{figure}

For a given proton decay mode, the energy and momentum distributions of final states only depend on those of decayed protons if we ignore the FSI and de-excitation processes of residual nucleus. The free proton from Hydrogen has an energy of $m_p=938.27$ MeV and the zero momentum. Compared to the free proton, the energy and momentum of bound protons in the Carbon nucleus will be influenced by the nuclear effects \cite{Super-Kamiokande:2014otb}, including the nuclear binding energy, the Fermi motion and the Nucleon-Nucleon correlations. In GENIE (version 3.0.2 with the G18\_02a\_00\_000 model set), the Bodek-Ritchie relativistic Fermi gas nuclear model \cite{Bodek:1981wr} with short-range NN correlation is used for all processes \cite{GENIE}. In this case, the off-shell energy $E_p$ of bound protons in $^{12} {\rm C}$ is given by
\begin{eqnarray} \label{eq1}
E_p = M_{^{12} {\rm C}} - \sqrt{ M^2_{^{11} {\rm B}} + k_1^2} \;,
\end{eqnarray}
where $M_{^{12} {\rm C}}$ ($M_{^{11} {\rm B}}$) denotes the ground-state nuclear mass of $^{12} {\rm C}$ ($^{11} {\rm B}$). In Fig. \ref{fig1}, we plot the momentum $k_1$ distribution of bound protons, where the tail $k_1 > 221$ MeV comes from the short-range NN correlation. By use of Eq. (\ref{eq1}), one can easily calculate $E_p$ as shown in the bottom panel of Fig. \ref{fig1}. It is found that $E_p$ has a very narrow range of $910.0 \, {\rm MeV} \leq E_p \leq 922.4$ MeV. Meanwhile, the $E_p$ shape doesn't display the nuclear shell structure, which is not consistent with the measured $s$-shell and $p$-shell binding energies from the $^{12} {\rm C}(e,e' p) ^{11} {\rm B}$ experiment \cite{Nakamura:1976mb}. In addition, the residual nucleus $^{11} {\rm B}$ associated with proton decays in $^{12} {\rm C}$ is assumed to be the ground state, which means $^{11} {\rm B}$ does not emit any particles through the de-excitation processes.

To solve the above two problems, the off-shell energy $E_p$ of bound protons in $^{12} {\rm C}$ is modified by
\begin{eqnarray} \label{eq2}
E_p = M_{^{12} {\rm C}} - \sqrt{ M^2_{^{11} {\rm B}^*} + k_1^2} \;,
\end{eqnarray}
where $M_{^{11} {\rm B}^*} = M_{^{11} {\rm B}} + E_x$ and $E_x$ is the excited energy of the residual nucleus $^{11} {\rm B}$. Note that Eq. (\ref{eq2}) is usually used in the analysis of electron/neutrino scattering from a nucleon bound in a nucleus \cite{Bodek:2018lmc, MINERvA:2019ope}. Due to $k_1 \ll M_{^{11} {\rm B}^*}$, we may simplify Eq. (\ref{eq2}) and derive
\begin{eqnarray} \label{eq3}
E_p \approx m_p - (S^p + E_x + \frac{k_1^2}{2 M_{^{11} {\rm B}}}) = m_p - E_R \;,
\end{eqnarray}
where $S^p =  M_{^{11} {\rm B}} + m_p - M_{^{12} {\rm C}} = 15.96$ MeV is the proton separation energy and $E_R = S^p + E_x + \frac{k_1^2}{2 M_{^{11} {\rm B}}}$ is defined as the removal energy \cite{Bodek:2018lmc, MINERvA:2019ope}. $E_R$ can also be expressed by $E_R = E_B + \frac{k_1^2}{2 M_{^{11} {\rm B}}}$ and $E_B = S^p + E_x$ is the binding energy. In order to describe the initial state of decayed protons, we adopt the Spectral Function (SF) nuclear model \cite{Benhar:2005dj}, which provides a 2-dimensional distribution of momentum $k_1$ and removal energy $E_R$ for bound protons in $^{12} {\rm C}$. Note that the SF model is also an optional nuclear model in GENIE. The SF nuclear model can well describe the neutrino-nucleus reactions \cite{MINERvA:2019ope,T2K:2018rnz,MINERvA:2018hba}, and could be further improved with continued experimental investigations. Based on the GENIE SF model and the modified $E_p$ in Eq. (\ref{eq3}), we plot the $k_1$ and $E_p$ distributions in Fig. \ref{fig1}. It is clear that the $E_p$ distribution shows the shell structure. The left and right peaks refer to the $s$-shell and $p$-shell protons, respectively.

\begin{figure}
\includegraphics[scale=0.4]{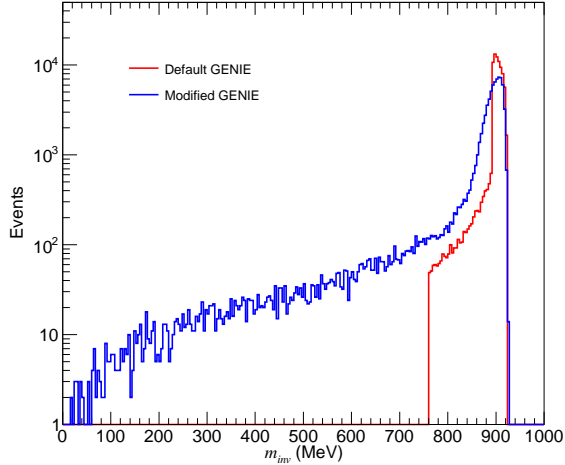}
\caption{Invariant mass distributions of bound protons from $^{12} {\rm C}$ in the default (red) and modified (blue) GENIE.}
\label{tM}
\end{figure}

With the help of $k_1$ and $E_p$, one can easily calculate the invariant mass $m_{inv}$ of bound protons in $^{12} {\rm C}$. Compared with the modified GENIE, the default GENIE has a more narrow range as shown in Fig. \ref{tM}. For the proton decay to be possible, the invariant mass of the initial proton has to be larger than the sum of masses of final products, namely $m_{inv} > \sum m_{f}$. Otherwise, this decay will be kinematically forbidden. Based on the invariant mass spectra in Fig. \ref{tM}, we calculate the forbidden ratios for all possible two-body proton decay modes, including an antilepton and a meson. The forbidden ratio is defined as the fraction of the initial-state protons for which a given decay mode is kinematically forbidden. As listed in Table \ref{forbidden-ratio}, the modified GENIE has a greater forbidden ratio than the default GENIE for every proton decay mode. This is because that $m_{inv}$ has a long tail below 760 MeV in the modified GENIE case. It is worthwhile to stress that significant differences between the default and modified GENIE can be found for some proton decay modes of $\sum m_{f}$ close to the proton mass.

\begin{table}
\centering
\caption{ Forbidden ratios of all possible two-body proton decay modes including an antilepton and a meson in the default and modified GENIE.}
\begin{tabular}{|c|c|c|c|} \hline  \hline
Decay mode       &    $\sum m_{f}$ (MeV)    &    \multicolumn{2}{c|}{Forbidden ratio ($\%$)}  \\  \hline
      &                   &    default        &  modified  \\ \hline
$p\rightarrow e^{+}\pi^0$            &       135.5       &      0     & 0.2        \\ \hline
$p\rightarrow \mu^{+}\pi^0$          &       240.6       &      0     & 0.5        \\ \hline
$p\rightarrow \bar{\nu} \pi^+$             &       139.6       &      0     & 0.2        \\ \hline
$p\rightarrow e^{+}\eta$             &       548.4       &      0     & 2.8        \\ \hline
$p\rightarrow \mu^{+}\eta$           &       653.5       &      0     & 4.3        \\ \hline
$p\rightarrow e^{+}\rho^0$           &       775.8       &      0.2   & 7.8        \\ \hline
$p\rightarrow \mu^{+}\rho^0$         &       880.9       &      5.4   & 28.3       \\ \hline
$p\rightarrow \bar{\nu} \rho^+$            &       763.0       &      0     & 7.3        \\ \hline
$p\rightarrow e^{+}\omega$           &       783.2       &      0.4   & 8.2        \\ \hline
$p\rightarrow \mu^{+}\omega$         &       888.3       &      6.5   & 37.8       \\ \hline
$p\rightarrow e^{+}K^0$              &       498.1       &      0     & 2.2        \\ \hline
$p\rightarrow \mu^{+}K^0$            &       603.3       &      0     & 3.5        \\ \hline
$p\rightarrow \bar{\nu} K^+$               &       493.7       &      0     & 2.2        \\ \hline
$p\rightarrow e^{+}K^{*}(892)^0$     &       896.1       &      21.5  & 51.2       \\ \hline
$p\rightarrow \bar{\nu} K^{*}(892)^+$      &       891.7       &      7.2   & 43.1       \\ \hline
  \hline
 \end{tabular}
\label{forbidden-ratio}
\end{table}


\section{\label{sec3} De-excitation Processes of Residual Nuclei}

As mentioned above, the residual nucleus $^{11} {\rm B}$ is always left in the ground state, and the de-excitation process is not taken into account in the default GENIE. In fact, the $s$-shell proton decay in $^{12} {\rm C}$ will leave a $s$-hole state of $^{11} {\rm B}$, which de-excites by emitting $n, p, d, t$, $\alpha$ and $h$ ($^{3} {\rm He}$) particles. The previous experiments have measured some two-body de-excitation modes for the excitation energy range of 16 MeV $\leq E_x \leq$ 35 MeV \cite{Yosoi:2003jjb,Panin:2016div}. For the higher $E_x$ region or three-body de-excitation modes, we do not know the corresponding branching ratio of every de-excitation mode from the $s$-hole state of $^{11} {\rm B}$. In addition to $^{11} {\rm B}$, the daughter particles of proton decays may interact with the spectator nucleons, and knock one or more of them out of $^{11} {\rm B}$. Due to the FSI, the residual nucleus $^{11} {\rm B}$ may be converted to $^{10} {\rm B}$ or $^{10} {\rm Be}$, and they are left in many particle-hole states. In order to determine the de-excitation processes of $^{11} {\rm B}$, $^{10} {\rm B}$ and $^{10} {\rm Be}$, the statistical model calculation is need. Here we shall calculate the excitation energy spectra of three kinds of residual nuclei based on the modified GENIE, and then use the TALYS code to estimate branching ratios of their de-excitation modes. TALYS is a nuclear reaction program, and is extensively used for both basic and applied science \cite{TALYS,Cheng:2020aaw,Cheng:2020oko}. In Sec. \ref{sec4}, we shall compare the predicted branching ratios with the experimental measurements \cite{Yosoi:2003jjb,Panin:2016div}.


\subsection{\label{sec3.1}  De-excitation processes of $^{11} {\rm B}$ states}

\begin{figure}
\includegraphics[scale=0.4]{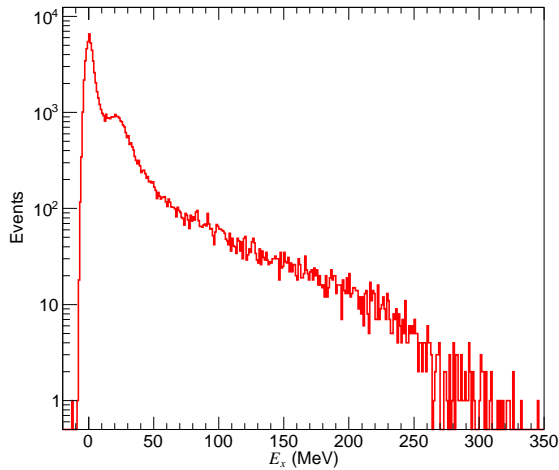}
\caption{Excitation energy $E_x$ distribution of the residual nucleus $^{11} {\rm B}$ after proton decays in $^{12} {\rm C}$.}
\label{Ex}
\end{figure}

In the modified GENIE, the excitation energy $E_x$ can be derived from $E_x = E_R -  S^p - \frac{k_1^2}{2 M_{^{11} {\rm B}}}$. On the other hand, one can also use the invariant mass and ground-state mass of the residual nucleus $^{11} {\rm B}$ to calculate $E_x$. As shown in Fig. \ref{Ex}, the $E_x$ spectrum of $^{11} {\rm B}$ states has two peaks. The left peak (0.0 MeV) and right peak (20.2 MeV) correspond to the $p$-shell and $s$-shell proton decays in $^{12} {\rm C}$, respectively. Due to the NN nuclear correlation, the excitation energy $E_x$ may be up to 300 MeV.

In a simple shell model of $^{12} {\rm C}$, four protons occupy the $p_{3/2}$ orbit, and two protons occupy the $s_{1/2}$ orbit. Due to pairing effects, a pair of protons may be in the $p_{1/2}$ shell in about $40 \%$ of the times \cite{Kamyshkov:2002wp}. In any case, $33.33 \%$ of $^{11} {\rm B}$ states are left in a highly excited state because of a $s_{1/2}$ proton decay. For the case of two protons in the $p_{1/2}$ shell, there is a one-third chance that the decayed protons come from the $p_{3/2}$ shell. Therefore,  $13.33 \%$ of $^{11} {\rm B}$ states are excited with medium excitation energy. For the other $53.33 \%$ possibility,  the residual nucleus $^{11} {\rm B}$ is assumed to be the ground state. Based on the $E_x$ distribution in Fig. \ref{Ex}, $53.33 \%$, $13.33 \%$, and $33.33 \%$ possibilities correspond to $E_x \leq 6.2$ MeV, $6.2 \; {\rm MeV} < E_x < 15.9$ MeV, and $E_x \geq 15.9$ MeV ranges, respectively. For simplicity, we require that $^{11} {\rm B}$ states with $E_x \leq 6.2 $ MeV do not de-excite, $^{11} {\rm B}$ states with $6.2 \; {\rm MeV} < E_x < 15.9$ MeV will emit a $\gamma$-ray with the energy of $E_x$.

\begin{table}
\centering
\caption{ The de-excitation modes, corresponding thresholds and branching ratios from the residual nucleus $^{11} {\rm B}$ in the case of a $s_{1/2}$ proton decay.}
\begin{tabular}{|c|c|c|} \hline  \hline
De-excitation Mode           & Threshold (MeV)  &    Branching ratio ($\%$)   \\ \hline
$n + {^{10}{\rm B}}$              &       11.5    &      8.8        \\ \hline
$p + {^{10}{\rm Be}}$             &       11.2    &      3.1        \\ \hline
$d + {^{9}{\rm Be}}$              &       15.8    &      3.9        \\ \hline
$t + {^{8}{\rm Be}}$              &       11.2    &      2.4        \\ \hline
$\alpha + {^{7}{\rm Li}}$         &       8.7     &      2.0        \\ \hline
$2n + {^{9}{\rm B}}$              &       19.9    &      3.5        \\ \hline
$n + p + {^{9}{\rm Be}}$          &       18.0    &      8.9        \\ \hline
$n+ d + {^{8}{\rm Be}}$           &       17.5    &      9.8        \\ \hline
$n + \alpha + {^{6}{\rm Li}}$     &       15.9    &     11.3        \\ \hline
$d + \alpha + {^{5}{\rm He}}$     &       18.1    &      1.1        \\ \hline
$t + 2\alpha $                    &       11.2    &      0.9        \\ \hline
$2n + p + {^{8}{\rm Be}}$         &       19.7    &      7.0        \\ \hline
$n + 2p + {^{8}{\rm Li}}$         &       34.9    &      0.6        \\ \hline
$2n + d + {^{7}{\rm Be}}$         &       36.4    &      0.5        \\ \hline
$n + p + d + {^{7}{\rm Li}}$      &       34.7    &      1.3        \\ \hline
$n + p + t + {^{6}{\rm Li}}$      &       35.7    &      0.7        \\ \hline
$2n + \alpha + {^{5}{\rm Li}}$    &       19.9    &      0.9        \\ \hline
$n + p + \alpha + {^{5}{\rm He}}$ &       20.3    &      1.9        \\ \hline
$2n + 2p + {^{7}{\rm Li}}$        &       37.0    &      0.7        \\ \hline
$2n + p + d + {^{6}{\rm Li}}$     &       42.0    &      0.6        \\ \hline
other modes                       &       -       &      30.2       \\ \hline
\hline
 \end{tabular}
\label{modes}
\end{table}

\begin{table}
\centering
\caption{ The percentage of the ground state and discrete states for every daughter nucleus in Table \ref{modes} from the de-excitation of ${^{11}{\rm B}}$ states. The definition of ratio $R_A$ can be found in the text.}
\begin{tabular}{|c|c|c|c|c|c|c||c|c|c|c|c|} \hline  \hline
Nucleus  & \,\;   g.s  \; \, & \; 1st \; & \;  2nd  \; &  \; 3rd  \; &  \; 4th \;  &  \; 5th \;  & \; $R_A$ \; \\ \hline
${^{10}{\rm B}}$  & 13.2  & 36.2   & 14.1   & 25.9  &  6.3   & 4.3  & 0.141  \\ \hline
${^{9}{\rm B}}$   & 36.7  & 18.9   & 11.4   & 12.1  & 12.7   & 8.1  & 0.263  \\ \hline
${^{10}{\rm Be}}$ & 42.3  & 16.0   &  6.9   &  7.4  & 20.1   & 7.4  & 0.183  \\ \hline
${^{9}{\rm Be}}$  & 39.3  & 21.7   & 11.0   &  9.9  &  8.3   & 9.8  & 0.303  \\ \hline
${^{8}{\rm Be}}$  & 54.4  & 38.2   &  2.2   &  1.8  &  1.7   & 1.8  & 0.893  \\ \hline
${^{7}{\rm Be}}$  & 37.9  & 34.4   &  9.1   &  8.1  &  7.3   & 3.2  & 0.175  \\ \hline
${^{8}{\rm Li}}$  & 33.5  & 33.3   & 13.6   & 19.6  & -      & -    & 0.069  \\ \hline
${^{7}{\rm Li}}$  & 41.6  & 34.5   & 14.4   &  9.6  & -      & -    & 0.216  \\ \hline
${^{6}{\rm Li}}$  & 71.6  & 11.8   &  5.9   &  8.2  & -      & -    & 0.902  \\ \hline
\hline
 \end{tabular}
\label{B11_percent}
\end{table}

For the highly excited $^{11} {\rm B}$ states, we use the TALYS (version 1.95) code \cite{TALYS} to estimate the de-excitation processes via the standard ``projectile 0'' configuration. The inputting excitation energy grid of the initial $^{11} {\rm B}$  population is based on the excitation energy spectrum of $E_x \geq$ 15.9 MeV in Fig. \ref{Ex}. It has a bin width of 2 MeV and follows the spin and parity $J^\pi=1/2^+$ format. Then the TALYS code can calculate the de-excitations of the compound nucleus $^{11} {\rm B}$ through the multiple emission and give a summary table of exclusive cross sections. By use of the initial population cross section, one can easily obtain their branching ratios for these exclusive de-excitation modes. Note that the  daughter nuclei include the contributions from both the ground state and all discrete states for every exclusive de-excitation mode. The excited daughter nucleus may decay a $\gamma$-ray into its ground state, or emit a massive particle into another nucleus. However the particle emission ratios are still unknown for some discrete states of daughter nuclei from the NNDC database \cite{NNDC}, such as the 6th discrete level of $^{10} {\rm Be}$ and $^{9} {\rm Be}$. So we set the number of included discrete energy levels to be 5 for $^{10} {\rm B}$, $^{9} {\rm B}$, $^{10} {\rm Be}$, $^{9} {\rm Be}$,$^{8} {\rm Be}$,$^{7} {\rm Be}$, be 3 for $^{8} {\rm B}$, $^{9} {\rm Li}$, $^{8} {\rm Li}$, $^{7} {\rm Li}$, $^{6} {\rm Li}$, and be 1 for other daughter nuclei in the following TALYS calculations.

\begin{figure}
\includegraphics[scale=0.4]{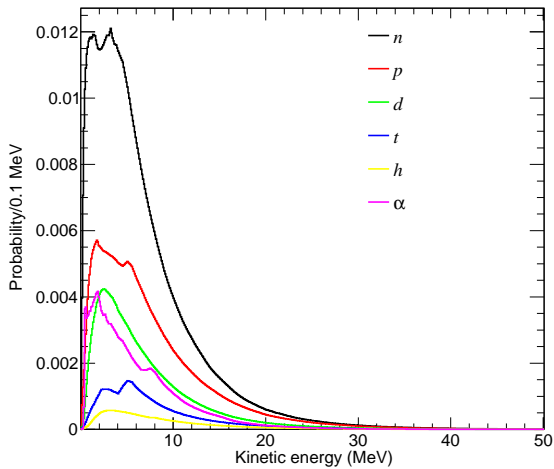}
 \caption{The kinetic energy distributions of six kinds of outgoing particles from the de-excitation processes of highly excited $^{11} {\rm B}$ states.}
\label{spectra_6}
\end{figure}

The predicted de-excitation modes, corresponding thresholds, and branching ratios of the highly excited $^{11} {\rm B}$ states are listed in Table \ref{modes}. The percentage of the ground state and discrete states for every daughter nucleus in Table \ref{modes} can be found in Table \ref{B11_percent}, those are calculated from their final populations after de-excitations of the continuous excited states of the corresponding nucleus. $R_A$ in the last column of Table \ref{B11_percent} describes the ratio that this nucleus will end up in the ground state and discrete states from its initial population after de-excitations of the continuous excited states in TALYS. Namely, the final population of the ground state and discrete states is $R_A$ of the initial population. The other population (1-$R_A$) is converted into the lighter daughter nuclei because of the emissions of massive particles from the continuous states. From Table \ref{modes}, one can find that the dominant de-excitation modes, including $n + {^{10} {\rm B}}$, $n + p +  {^{9} {\rm Be}}$, $ n + d + {^{8} {\rm Be}}$, $n + \alpha +   {^{6} {\rm Li}}$, $2n +p + {^{8} {\rm Be}}$, will contribute at a branching ratio of $45.8\%$ for a $s_{1/2}$ proton decay. About $56.5 \%$ of $^{11} {\rm B}$ states can directly produce one or more neutrons from the listed exclusive modes. All exclusive de-excitation modes with a branching ratio of less than $0.5\%$ and non-exclusive de-excitation processes have been classified as the other modes. It is found that the multiplicities of $n, p, d, t, h, \alpha$ emissions are 1.35, 1.02, 0.54, 0.30, 0.18 and 0.35 for the other modes, respectively. The other modes mainly originate from the $^{11} {\rm B}$ states with $E_x >50$ MeV since the excitation energy is enough to break the $^{11} {\rm B}$ nucleus into pieces. In Fig. \ref{spectra_6}, we plot the kinetic energy distributions of six kinds of outgoing particles from both the exclusive modes and the other modes. Then one can obtain the average kinetic energies 6.6, 7.7, 7.0, 8.0, 8.5 and 6.1 MeV for $n, p, d, t, h$ and $\alpha$ particles, respectively. Based on the neutron kinetic energy distribution in Fig. \ref{spectra_6}, most of them will give a 2.2 MeV $\gamma$-ray from the neutron capture reaction in LS detectors. In addition, the de-excitation modes of $d +  {^{9} {\rm Be}}$ and $d + \alpha +  {^{5} {\rm He}}$ can indirectly produce a neutron from the neutron emission of five discrete levels of ${^{9} {\rm Be}}$ and the ${^{5} {\rm He}}$ decay.

\subsection{\label{sec3.2} De-excitation processes of $^{10} {\rm B}$ and $^{10} {\rm Be}$ states}

In the above subsection, we only consider the de-excitation processes of the residual nucleus $^{11} {\rm B}$. Other residual nuclei can also be produced through the FSI. Due to the strong interaction, the daughter mesons from the proton decay may interact with spectator nucleons before escaping from the residual nucleus surface. In this case, the FSI can knock the struck nucleon out of $^{11} {\rm B}$, and the residual nucleus is converted to the $^{10} {\rm B}$ or $^{10} {\rm Be}$ nucleus. The multiple FSI can result in other residual nuclei, such as $^{9} {\rm B}$, $^{9} {\rm Be}$, and $^{8} {\rm Li}$, etc. Here we focus on the de-excitation processes of $^{10} {\rm B}$ and $^{10} {\rm Be}$ states.

\begin{figure}
\includegraphics[scale=0.4]{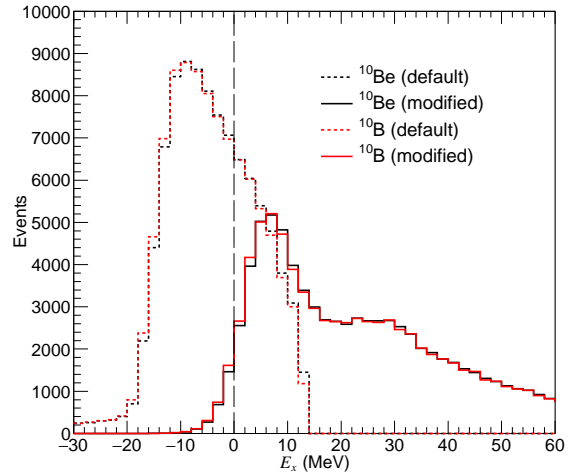}
\caption{Excitation energy spectra of $^{10} {\rm B}$ and $^{10} {\rm Be}$ in the default and modified GENIE.}
\label{Nk}
\end{figure}

In the default GENIE, the spectator nucleon is on-shell and its energy is given by
\begin{eqnarray} \label{eq4}
E_N = \sqrt{m^2_N + k_2^2} \;,
\end{eqnarray}
where $m_N$ is the nucleon mass and the momentum $k_2$ is same with the red line in the top panel of Fig. \ref{fig1}. With the help of Eqs. (\ref{eq1}) and (\ref{eq4}), one can calculate the invariant masses of $^{10} {\rm B}$ and $^{10} {\rm Be}$ through
\begin{eqnarray} \label{eq5}
M_{inv} = \sqrt{(M_{^{12} {\rm C}}-E_p-E_N+ E_b)^2 - (\vec{k}_1 + \vec{k}_2)^2} \;.
\end{eqnarray}
Since Eq.(\ref{eq4}) overestimates the energy of spectator nucleon, the default GENIE uses a fixed binding energy $E_b =25$ MeV to offset the reduction of residual nucleus energy. Then the excitation energy spectra of $^{10} {\rm B}$ and $^{10} {\rm Be}$ states can be easily obtained from $E_x = M_{inv} - M_R$ as shown in Fig. \ref{Nk}, where $M_R$ is the corresponding mass in the ground state. It is clear that the average excitation energy $\langle E_x \rangle$ is negative in the default GENIE case.

\begin{table}
\centering
\caption{ De-excitation modes, corresponding thresholds and branching ratios from the de-excitation of $^{10} {\rm B}$ states.}
\begin{tabular}{|c|c|c|} \hline  \hline
De-excitation Mode           & Threshold (MeV)  &    Branching ratio ($\%$)   \\ \hline
$\gamma + {^{10}{\rm B}}$         &       0       &      12.2         \\ \hline
$n + {^{9}{\rm B}}$               &       8.4     &      4.9        \\ \hline
$p + {^{9}{\rm Be}}$              &       6.6     &      7.1       \\ \hline
$d + {^{8}{\rm Be}}$              &       6.0     &      8.0        \\ \hline
$t + {^{7}{\rm Be}}$              &       18.7    &      0.6        \\ \hline
$h + {^{7}{\rm Li}}$              &       17.8    &      0.6        \\ \hline
$\alpha + {^{6}{\rm Li}}$         &       4.5     &     10.3        \\ \hline
$2n + {^{8}{\rm B}}$              &       27.0    &      0.6        \\ \hline
$n + p + {^{8}{\rm Be}}$          &       8.3     &     11.2        \\ \hline
$2p + {^{8}{\rm Li}}$             &       23.5    &       0.8        \\ \hline
$n + d + {^{7}{\rm Be}}$          &       24.9    &      1.6       \\ \hline
$p + d + {^{7}{\rm Li}}$          &       23.3    &      1.7       \\ \hline
$p + t + {^{6}{\rm Li}}$          &       24.3    &      1.1       \\ \hline
$n + h + {^{6}{\rm Li}}$          &       25.0    &      1.1       \\ \hline
$n + \alpha + {^{5}{\rm Li}}$     &       10.1    &      1.9       \\ \hline
$p + \alpha + {^{5}{\rm He}}$     &       8.9     &      2.3       \\ \hline
$2n + p + {^{7}{\rm Be}}$         &       27.2    &      0.6        \\ \hline
$n + 2p + {^{7}{\rm Li}}$         &       25.5    &      1.0        \\ \hline
$n + p + d + {^{6}{\rm Li}}$      &       30.5    &      1.0        \\ \hline
$p + d + t + \alpha       $       &       25.7    &      0.5        \\ \hline
$2n + 2p + {^{6}{\rm Li}} + \alpha $ &      26.5    &      0.5        \\ \hline
other modes                            &       -      &       30.6       \\ \hline
 \hline
 \end{tabular}
\label{B10}
\end{table}

\begin{table}
\centering
\caption{ The percentage of the ground state and discrete states for every daughter nucleus in Table \ref{B10} from the de-excitation of ${^{10}{\rm B}}$ states. The definition of ratio $R_A$ can be found in Sec. \ref{sec3.1}.}
\begin{tabular}{|c|c|c|c|c|c|c||c|c|c|c|c|} \hline  \hline
Nucleus  & \,\;   g.s  \; \, & \; 1st \; & \;  2nd  \; &  \; 3rd  \; &  \; 4th \;  &  \; 5th \;  & \; $R_A$ \;  \\ \hline
${^{10}{\rm B}}$  & 22.4  & 13.9   & 10.6  & 16.6  & 23.2   & 13.4   & 0.097  \\ \hline
${^{9}{\rm B}}$   & 33.3  & 17.9   & 13.6  & 12.8  & 12.9   & 9.5    & 0.169 \\ \hline
${^{8}{\rm B}}$   & 34.7  & 33.2   & 14.6  & 17.5  & -      &  -     & 0.078  \\ \hline
${^{9}{\rm Be}}$  & 41.3  & 17.4   & 11.6  & 11.2  & 10.2   & 8.3    & 0.198 \\ \hline
${^{8}{\rm Be}}$  & 47.0  & 39.9   &  4.0  &  3.2  &  3.0   & 2.8    & 0.827 \\ \hline
${^{7}{\rm Be}}$  & 36.5  & 34.2   & 10.0  &  8.3  &  7.5   & 3.5    & 0.215 \\ \hline
${^{8}{\rm Li}}$  & 34.8  & 31.7   & 14.3  & 19.2  & -      & -      & 0.084  \\ \hline
${^{7}{\rm Li}}$  & 46.7  & 35.0   &  9.9  &  8.4  & -      & -      & 0.213 \\ \hline
${^{6}{\rm Li}}$  & 66.6  & 16.2   &  6.6  &  1.1  & -      & -      & 0.844 \\ \hline
\hline
 \end{tabular}
\label{B10_percent}
\end{table}

In order to solve the $\langle E_x \rangle < 0$ problem, we modify the spectator nucleon energy with
\begin{eqnarray} \label{eq6}
E_N = m_N - E_R \;,
\end{eqnarray}
where the removal energy $E_R$ is given by the SF nuclear model \cite{Benhar:2005dj}. Due to the absence of the $^{11} {\rm B}$ spectral function in GENIE, we assume spectator nucleons in $^{11} {\rm B}$ have the same SF distribution with protons in $^{12} {\rm C}$. In this case, the invariant masses of the residual nuclei $^{10} {\rm B}$ and $^{10} {\rm Be}$ can be derived from
\begin{eqnarray} \label{eq7}
M_{inv} = \sqrt{(M_{^{12} {\rm C}}-E_p-E_N)^2 - (\vec{k}_1 + \vec{k}_2)^2} \;.
\end{eqnarray}
Using $E_x = M_{inv} - M_R$, we plot the $E_x$ spectra of $^{10} {\rm B}$ and $^{10} {\rm Be}$ states in Fig. \ref{Nk}. The difference between the two curves is very small.

About $2.7 \%$ $^{10} {\rm B}$ or $^{10} {\rm Be}$ states will be assumed to be the ground state because of $E_x < 0$. Then we use the TALYS code to estimate the de-excitation processes of $^{10} {\rm B}$ and $^{10} {\rm Be}$ states based on the $E_x > 0$ distributions. The initial population is equally distributed over both parities and the inputting spin is assumed to zero. The branching ratio of each exclusive de-excitation mode from the residual nuclei $^{10} {\rm B}$ and $^{10} {\rm Be}$ have been listed in Table \ref{B10} and Table \ref{Be10}, respectively. The corresponding percentage of the ground state and discrete states for every daughter nucleus can be found in Tables \ref{B10_percent} and \ref{Be10_percent}. About $24.4 \%$ $^{10} {\rm B}$ and $45.3 \%$ $^{10} {\rm Be}$ states can directly produce one or more neutrons from the listed exclusive de-excitation modes. For the other modes of $^{10} {\rm B}$ ($^{10} {\rm Be}$) states, the multiplicities of $n, p, d, t, h, \alpha$ emissions are 1.00, 1.03, 0.42, 0.24, 0.21 and 0.30 (1.15, 0.62, 0.39, 0.29, 0.10 and 0.40), respectively.

\begin{table}
\centering
\caption{De-excitation modes, corresponding thresholds and branching ratios from the de-excitation of $^{10} {\rm Be}$ states.}
\begin{tabular}{|c|c|c|} \hline  \hline
De-excitation Mode           & Threshold (MeV)  &    Branching ratio ($\%$)   \\ \hline
$\gamma + {^{10}{\rm Be}}$        &       0       &      16.3      \\ \hline
$n + {^{9}{\rm Be}}$              &       6.8     &      19.6      \\ \hline
$d + {^{8}{\rm Li}}$              &       21.5    &      0.7        \\ \hline
$t + {^{7}{\rm Li}}$              &       17.2    &      1.2        \\ \hline
$\alpha + {^{6}{\rm He}}$         &       7.4     &      1.2        \\ \hline
$2n + {^{8}{\rm Be}}$             &       8.5     &     10.3        \\ \hline
$n + p + {^{8}{\rm Li}}$          &       23.7    &      1.7       \\ \hline
$n + d + {^{7}{\rm Li}}$          &       23.5    &      3.4       \\ \hline
$n + t + {^{6}{\rm Li}}$          &       24.5    &      1.8       \\ \hline
$n + \alpha + {^{5}{\rm He}}$     &       9.1     &      4.4       \\ \hline
$2n + p + {^{7}{\rm Li}}$         &       25.7    &      1.9       \\ \hline
$2n + d + {^{6}{\rm Li}}$         &       32.2    &      0.8       \\ \hline
$n + p + t + {^{5}{\rm He}}$      &       28.9    &      0.6        \\ \hline
$n + d + t + \alpha $             &       26.0    &      0.8        \\ \hline
other modes                       &       -       &      35.4       \\ \hline
  \hline
 \end{tabular}
\label{Be10}
\end{table}

\begin{table}
\centering
\caption{ The percentage of the ground state and discrete states for every daughter nucleus in Table \ref{Be10} from the de-excitation of ${^{10}{\rm Be}}$ states. The definition of ratio $R_A$ can be found in Sec. \ref{sec3.1}.}
\begin{tabular}{|c|c|c|c|c|c|c||c|c|c|c|c|} \hline  \hline
Nucleus  & \,\;   g.s  \; \, & \; 1st \; & \;  2nd  \; &  \; 3rd  \; &  \; 4th \;  &  \; 5th \; &  \; $R_A$ \;  \\ \hline
${^{10}{\rm Be}}$ & 31.7  & 41.3   & 20.0   &  2.9  & 2.4   & 1.7   & 0.140 \\ \hline
${^{9}{\rm Be}}$  & 48.0  & 16.0   & 10.8   & 10.1  & 8.5   & 6.6   & 0.306 \\ \hline
${^{8}{\rm Be}}$  & 38.7  & 47.5   &  3.8   &  3.5  & 3.3   & 3.2   & 0.848 \\ \hline
${^{8}{\rm Li}}$  & 36.1  & 31.0   & 15.2   & 17.7  & -     & -     & 0.095  \\ \hline
${^{7}{\rm Li}}$  & 48.0  & 35.9   &  8.7   &  7.4  & -     & -     & 0.239  \\ \hline
${^{6}{\rm Li}}$  & 47.4  & 21.4   & 13.0   & 18.2  & -     & -     & 0.600  \\ \hline
\hline
 \end{tabular}
\label{Be10_percent}
\end{table}


\section{\label{sec4} Comparison with experimental data}

\begin{figure}
\includegraphics[scale=0.4]{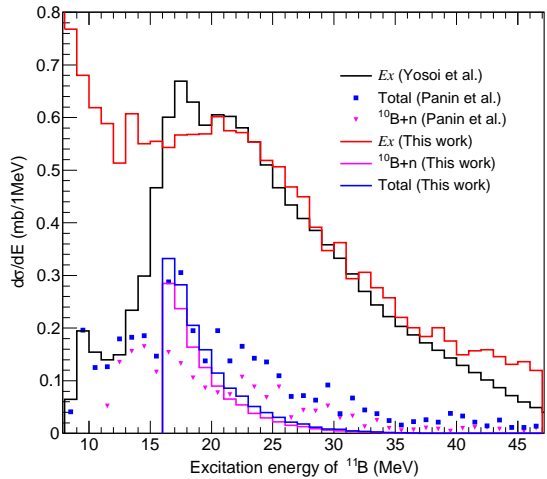}
\caption{Comparison of the measured (black line) \cite{Yosoi:2003jjb} and predicted (red line) excitation energy spectra of the residual nucleus $^{11} {\rm B}$. The pink triangle (pink line) and blue square (blue line) denote the measured (predicted) results \cite{Panin:2016div} for $^{11} {\rm B} \rightarrow n + ^{10} {\rm B}$ and all three two-body de-excitation modes, respectively.}
\label{Ex_com3}
\end{figure}

In order to study the properties of proton-hole states in $^{11} {\rm B}$, the quasifree $^{12} {\rm C} (p,2p) ^{11} {\rm B}$ \cite{Yosoi:2003jjb,Panin:2016div} and $^{12} {\rm C} (e,e'p) ^{11} {\rm B}$ \cite{Nakamura:1976mb} reactions have been extensively investigated. Yosoi et al. measured the excitation energy spectrum of the proton-hole states in $^{11} {\rm B}$ through the quasifree $^{12} {\rm C} (p,2p)  ^{11} {\rm B}$ reaction \cite{Yosoi:2003jjb}. As shown in Fig. \ref{Ex_com3}, the predicted $E_x$ curve (red line) from Fig. \ref{Ex} can well describe the measured one (black line) in the region of 15 MeV $\leq E_x \leq $ 40 MeV. Note that the measured $E_x$ spectrum from Ref. \cite{Yosoi:2003jjb} has been binned and scaled by an arbitrary factor to match the peak around 10 MeV in Fig. 4 of Ref. \cite{Panin:2016div}. For the convenience of following discussions, we have also scaled the predicted $E_x$ spectrum to match the scaled Yosoi spectrum. For $E_x < 15$ MeV, the predicted spectrum is inconsistent with the experimental measurement, and displays a continuous distribution as shown in Fig. \ref{Ex}. It can not explain several discrete peaks in Fig. 1a of Ref. \cite{Yosoi:2003jjb}. This is because that we use the continuous spectral function to calculate the excitation energy.

\begin{figure}
\includegraphics[scale=0.4]{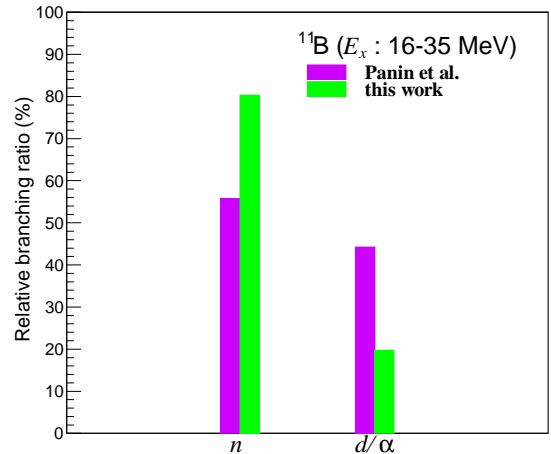}
\caption{Comparison of the measured (pink) and predicted (green) relative branching ratios of the $n$ and $d/\alpha$ emissions among all three two-body de-excitation modes from the residual nucleus $^{11} {\rm B}$ with 16 MeV $\leq E_x \leq $ 35 MeV. }
\label{ratio_Panin}
\end{figure}

In Ref. \cite{Panin:2016div}, Panin et al. measured three de-excitation modes of $^{11} {\rm B}$ states from the $^{12} {\rm C} (p,2p) ^{11} {\rm B}$ reaction, namely $^{11} {\rm B} \rightarrow n + ^{10} {\rm B}$, $^{11} {\rm B} \rightarrow d + ^{9} {\rm Be}$ and $^{11} {\rm B} \rightarrow \alpha + ^{7} {\rm Li}$. The $^{11} {\rm B} \rightarrow n + ^{10} {\rm B}$ (pink triangle) and total (blue square) spectra have been plotted in Fig. \ref{Ex_com3}. For comparison, we use the TALYS code to calculate branching ratios of the three de-excitation modes for every excitation energy bin in the range of 16-47 MeV. Besides the ground state, the predicted branching ratios includes the contributions from the first four discrete levels of $^{10} {\rm B}$ states, and the first excited level of $^{7} {\rm Li}$ states. In terms of the predicted $E_x$ distribution, we can derive the $^{11} {\rm B} \rightarrow n + ^{10} {\rm B}$ (pink line) and total (blue line) spectra as shown in Fig. \ref{Ex_com3}. It is clear that the predicted shapes of both the $n + ^{10} {\rm B}$ and total spectrum can not describe the measured shapes well. The experimental data of three two-body de-excitation modes in Fig. \ref{Ex_com3} can be converted into two relative branching ratios of the $n$ and $d/\alpha$ emissions among all three two-body de-excitation modes. The relative branching ratio of $n$ emission can be calculated by dividing the integral of the pink data points in Fig. \ref{Ex_com3} by the integral of the blue data points. Using the difference between the integrals of the blue and pink data points, one can also derive the relative branching ratio of $d/\alpha$ emission. Note that they are independent of the scale factor. In Fig. \ref{ratio_Panin}, we compare the predicted and measured relative branching ratios of the $n$ and $d/\alpha$ emissions for 16 MeV $ \leq E_x \leq $ 35 MeV. It is found that the predicted $n$ relative branching ratio is larger than the measured one. For the $d/\alpha$ emission, the predicted value can not account for the measured result.

\begin{figure}
\includegraphics[scale=0.4]{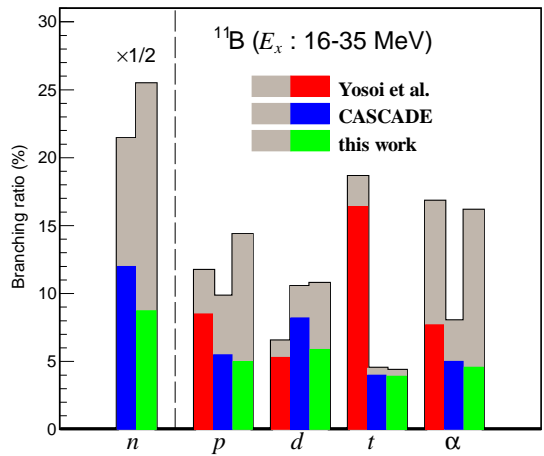}
\caption{Comparison of measured and predicted branching ratios of $n, p, d, t$ and $\alpha$ emissions from the residual nucleus $^{11} {\rm B}$ with 16 MeV $\leq E_x \leq $ 35 MeV. The experimental data (red) were taken from Ref. \cite{Yosoi:2003jjb}. The green and blue branching ratios denote the predicted results from the TALYS and CASCADE calculations, respectively.}
\label{ratio_Yosoi}
\end{figure}

In addition to the excitation energy spectrum, Ref. \cite{Yosoi:2003jjb} has also measured the branching ratios of charged particle emissions from the $s-$hole state in $^{11} {\rm B}$ as shown in Fig. \ref{ratio_Yosoi}. The colored areas indicate the branching ratios of decay onto the `two-body decay' regions, and the grey parts mainly come from contributions of three-body and sequential de-excitation modes of $^{11} {\rm B}$ states. In Ref. \cite{Yosoi:2003jjb}, the `two-body decay' regions are defined for each emitted particle species as the excitation energy range in the daughter nucleus between the ground state and the larger of 5 MeV and the threshold for emission of additional particles. The authors have provided simultaneously the results of a statistical model calculation from the CASCADE code \cite{Puhlhofer:1977zz}. Based on the same definition, we use the TALYS code to calculate the branching ratios of $n,p,d,t, \alpha$ emissions for every $E_x$ bin in the range of 16-35 MeV. Their detection threshold energies are 3.1, 3.1, 4.0, 4.6 and 4.5 MeV, respectively. In addition, the energetic $\alpha$ contribution from the decay of discrete levels of the daughter nucleus $^{8} {\rm Be}$ has been considered. With the help of the $E_x$ distribution between 16-35 MeV, we estimate the branching ratios of $n,p,d,t$ and $\alpha$ emissions from the residual nucleus $^{11} {\rm B}$. In Fig. \ref{ratio_Yosoi}, the green and blue branching ratios denote the TALYS and CASCADE results, respectively. It is found that the predicted branching ratio of each particle emission is basically consistent with the experimental data except for the $t$ emission. The TALYS and CASCADE codes derive similar results for all `two-body decay' modes. For the key $n$ emission, TALYS give the larger total branching ratio than CASCADE.

\section{\label{sec5} Summary}

In summary, we have investigated the de-excitation processes of the residual nuclei associated with proton decays in $^{12} {\rm C}$ based on the GENIE generator and TALYS code. Since the residual nucleus $^{11} {\rm B}$ would be left in the ground state, the default GENIE nucleon decay generator has been modified in terms of the SF nuclear model. After deriving the excitation energy spectra of $^{11} {\rm B}$, $^{10} {\rm B}$ and $^{10} {\rm Be}$, we use the TALYS code to implement their de-excitation processes. For the $s$-hole state in $^{11} {\rm B}$, we compare the predicted results with the experimental measurements. It is found that the predicted results are partly consistent with the experimental data. For $p, d$ and $\alpha$ emissions, the predicted branching ratios can basically account for the Yosoi measurement. The TALYS and CASCADE calculations give the similar results for all `two-body decay' modes. For the $n$ emission, the predicted shape from $^{11} {\rm B} \rightarrow n + ^{10} {\rm B}$ can not describe the measured shape well. For LS experiments, the total $n$ branching ratio from the residual nucleus $^{11} {\rm B}$ is the most crucial quantity due to the neutron capture. However, it has not been measured up to now. We hope that future experiments can measure the total branching ratio including one or more neutrons from de-excitation processes of the residual nucleus $^{11} {\rm B}$.

\acknowledgments

We are grateful to Jie Cheng, Xianguo Lu, Yufeng Li, Yuhang Guo, Benda Xu and Aiqiang Zhang for their helpful discussions. This work is supported in part by the National Nature Science Foundation of China (NSFC) under Grants No. 11575201, No. 11675273 and No. 11775316, and the Strategic Priority Research Program of the Chinese Academy of Sciences under Grant No. XDA10010100.

\end{document}